\documentclass[%
 reprint,
superscriptaddress,
 amsmath,amssymb,
 aps,
]{revtex4-2}

\usepackage{graphicx}
\usepackage{dcolumn}
\usepackage{bm}
\usepackage{physics}
\usepackage[dvipsnames]{xcolor}
\usepackage[
	colorlinks=true,
	citecolor=Red,
	linkcolor=Blue,
	filecolor=magenta,
	urlcolor=Blue,
]{hyperref}

\newcommand{\ITMO}{School of Physics and Engineering, ITMO University,
197101 St. Petersburg, Russia} 
\newcommand{\Qingdao}{Qingdao Innovation and Development Center, Harbin Engineering University, Qingdao 266000, Shandong, China}
\newcommand{\Tongji}{Institute of Acoustics, School of Physics Science and Engineering, Tongji University, Shanghai 200092, China}
\newcommand{\iu}{{i}\mkern1mu}
\newcommand{\eu}{\mathrm{e}\mkern1mu}

\begin{document}


\preprint{AIP/123-QED}

\title[Experimental investigation of Acoustic Kerker Effect in Labyrinthine Resonators]{Experimental investigation of Acoustic Kerker Effect in Labyrinthine Resonators}
\author{Iuliia Timankova}
\affiliation{\ITMO}

\author{Mikhail Smagin}
\affiliation{\ITMO}

\author{Mikhail Kuzmin}
\affiliation{\ITMO}

\author{\\Andrey Lutovinov}
\affiliation{\ITMO}

\author{Andrey Bogdanov}
\affiliation{\Qingdao}
\affiliation{\ITMO}
\author{Yong Li}
\affiliation{\Tongji}
\affiliation{\ITMO}%

\author{Mihail Petrov}
\affiliation{\ITMO}%
 \email{m.petrov@metalab.ifmo.ru}


\begin{abstract}
Controlling the directionality of the acoustic scattering with single acoustic metaatoms has a key importance for reaching spatial routing of sound with acoustic metamaterials.  In this paper, we present the experimental demonstration of the acoustic analogue of the Kerker effect realized in a two-dimensional coiled-space metaatom. By engineering the interference between monopolar and dipolar resonances within a high-index acoustic metaatom, we achieve directional scattering with suppressed backward or forward response at the first and second Kerker conditions respectively. Experimental measurements of the scattered pressure field, in a parallel-plate waveguide environment, show good agreement with the full-wave simulations. Our results validate the feasibility of Kerker-inspired wave control in acoustic systems and open new opportunities for directional sound manipulation.
\end{abstract}

\maketitle

The study of directional light scattering has gained significant attention in nanophotonics over the past decade~\cite{liu2018}, owing to its relevance for wave control phenomena such as perfect transmission, reflection, and absorption, as well as for the field of optical particle manipulation~\cite{gao2017}. The fundamental mechanism behind directional scattering is the Kerker effect~\cite{kerker1983}, which arises from the interference between electric and magnetic dipole contributions in the scattered field. This type of interference can be realized in high refractive-index dielectric particles and in other resonant photonic platforms supporting strong multipole responses~\cite{poshakinskiy2019,gerasimov2021}. The framework of multipole analysis, initially developed as a rigorous approach to describe scattering processes, has since evolved into the broader concept of multipole engineering~\cite{liu2020nanophotonics}, now central to the design of tailored optical responses. More recently, analogous ideas have been extended to acoustics, where multipole decomposition has emerged as a powerful methodology~\cite{baresch2016, tsimokha2022}, for analyzing directional scattering in particular~\cite{wei2020,wu2021,Smagin2024, toftul2024arxiv,toftul2025acoustic}. In this letter, we consider the acoustic analogy of a Kerker effect. Despite these recent theoretical and numerical demonstrations, direct experimental realization of the acoustic analogue of the Kerker effect, well established in optics since its first experimental observation in 2012~\cite{geffrin2012}, has not been reported so far. 

\begin{figure}[t!]
\centering
\includegraphics[width=0.97\linewidth]{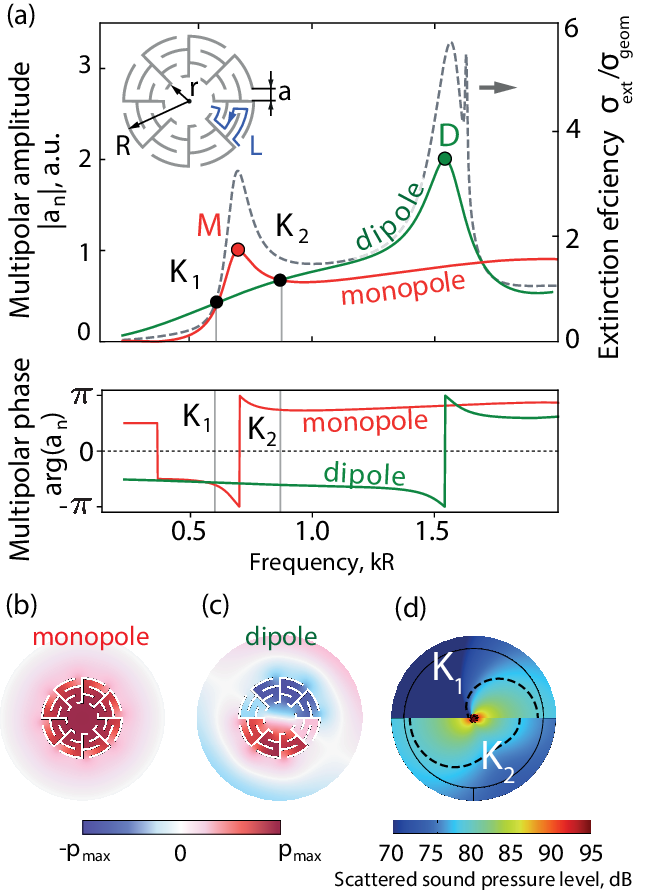}
\caption{(a) Cylindrical function expansion of the field scattered by labyrinthine metaatom with geometry parameters $R=2.5$~cm, $r=0.9$~cm, $a=4$~mm, 8~ sectors, and a curling number of 3. The points where first and second Kerker conditions are satisfied are marked with $K_1$ and $K_2$ respectively. (b) Monopole and (c) Dipole modes in a high-index labyrinthine metaatom. (d) Radiation patterns at Kerker and anti-Kerker conditions.  }
\label{fig1:concept}
\end{figure}

In its original formulation, the Kerker effect corresponds to the interference between electric and magnetic dipolar radiation, which possess opposite parities in the far field~\cite{kerker1983}. When these two contributions are of equal magnitude and in phase, the backward component of the scattered wave is suppressed, whereas an out-of-phase relation leads to the cancellation of forward scattering instead. A similiar scenario can be exploited in acoustics, where two lowest order multipoles, monopole and dipole, have the same dependence on the characteristic size of the particle \cite{toftul2019}, but differ in parity~\cite{Sun2020}. In view of this analogy, recent theoretical work~\cite{wei2020} has introduced an acoustic counterpart of the Kerker effect, achieved through the interference between monopolar and dipolar resonances of a high-refractive-index scatterer.

\begin{figure}[t!]
\centering
\includegraphics[width=0.95\linewidth]{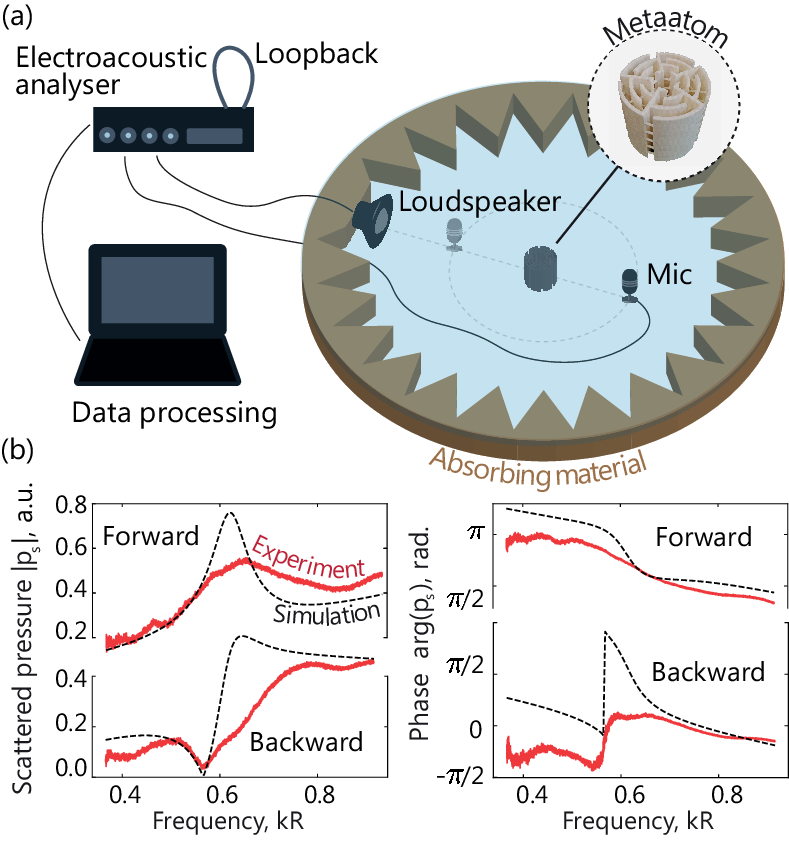}
\caption{(a) Schematic of the experimental setup for two-dimensional acoustic scattering measurements. The system comprises a parallel-plate waveguide constructed from two thin plexiglass sheets, with the edges terminated by wedges of sound-absorbing rubber foam to suppress boundary reflections. The overall dimensions of the setup are $1.5 \times 1.5 \times 0.066$~m, and the radiation patterns are recorded at a radial distance of $15$~cm. The signal is a chirped sweep of all covered frequencies. The signals are synchronized using loopback, and recorded using the sound card for further processing. (b) Amplitude and phase of the scattered at forwards and backwards points of the metaatom
  }
\label{fig2:setup_data}
\end{figure}
Unlike in optics, naturally occurring high-index particles are absent in airborne acoustics. However, artificial structures with labyrinthine geometries can emulate this behavior by substantially elongating the effective propagation path of sound waves~\cite{cheng2015,zhang2022}, thereby reducing the effective sound velocity and enabling resonators of subwavelength dimensions. Such coiled-space metaatoms serve as acoustic analogues of high-index dielectric particles, which makes them an attractive platform to study resonant multipole effects. Here, we utilize a two-dimensional implementation of such particle. Its cross section is shown in Fig.~\ref{fig1:concept}(a) inset. In such particles, the refractive index relative to the surrounding medium is determined by the structural parameters of the labyrinth and can be written as $n = L/(R_2-R_1)$~\cite{Zhu2016}, where $R_2$ and $R_1$ denote the outer and inner radii of the structure, respectively, and $L$ is the effective propagation length of sound within the structure. The resonant acoustic features of labyrinth metaatoms  can be clearly seen in the plane wave scattering spectrum shown in Fig.~\ref{fig1:concept}(a) showing peaks corresponding to the monopole and dipole resonances.

For two dimensional geometry with translational invariance along the axis perpendicular to the figure plane, the local multipole moments are directly connected to cylindrical harmonic expansion of the scattered field \cite{blackstock2000} written as:
\begin{equation}
\label{eq:cylindrical_expansion}
    p_s = \sum \limits_{n}  A_n H_n^{(1)}(kr) \eu^{\iu n\theta},
\end{equation}
where $n$ is the multipole order, $A_n$ are expansion coefficients of the scattered field, $H_n^{(1)}$ are Hankel functions of the first kind. Here and further we assume all fields to have harmonic time-dependence, $\bar{p}(\vb{r}, t) = \Re [p(\vb{r}) \eu^{- \iu \omega t}]$. The calculated partial extinction cross sections corresponding to different multipolar terms are also shown in Fig.~\ref{fig1:concept} (a) with dashed lines demonstrating strong monopole and dipole contributions for the first and the second resonance. The monopole and dipole character of the mode is also clearly seen in Figs.~\ref{fig1:concept}(b) and \ref{fig1:concept}(c) demonstrating pressure field distribution of the corresponding eigenmodes.

The requirements for directional scattering can be formulated through the relation between the monopolar and dipolar coefficients: constructive interference in the forward direction is obtained when $A_0 = A_1$, while cancellation of the forward lobe (i.e., enhanced backscattering) occurs for $A_0 = -A_1$. Numerical studies have shown that these conditions can indeed be fulfilled in the considered geometry~\cite{wu2021} at two wavelengths which we denote as $K_1$ and $K_2$ in Fig.~\ref{fig1:concept}(a). The far field distribution along with modeled  far field scattering patterns are shown in Figs.~\ref{fig1:concept}(d) and \ref{fig1:concept}(e).

To observe the discussed effect, a parallel-plate waveguide geometry was used [see the setup scheme in Fig.~\ref{fig2:setup_data} (a)]. Such waveguide supports a single propagating mode with homogeneous field distribution along the $z$-axis. In our experimental setup the waveguide was formed of two plexiglass sheets with the separation of $6.6$~cm, which ensures the single mode  propagation up to the frequency of  about $2500$~Hz, thereby creating an effectively two-dimensional acoustic environment. The size of the chamber is $1.5 \,\text{m} \times 1.5 \,\text{m}$ and parasitic reflections at the boundaries of chamber were suppressed by adding absorbing material and wedge-shaped terminations. 
The operational frequency window was chosen to be from 500~Hz to 2500~Hz. The lower limit arises from the reduced absorption efficiency of the boundary treatment at longer wavelengths, while the upper limit is dictated by the cutoff of higher-order modes. To design the geometry of the fabricated metaatom [see Fig.~\ref{fig1:concept}], full-wave numerical simulations were conducted in \textsc{Comsol Multiphysics}. 

The metaatom used in the measurements, shown in the inset of Fig.~\ref{fig2:setup_data}(a), was fabricated via additive manufacturing (FDM 3D printing) using PLA polymer. The chosen geometrical sizes  provide  the first and second Kerker conditions  at the frequency of around 800~Hz and 1200~Hz, respectively.   The real 3D metaatom structure was additionally reinforced with plastic caps at the top and the bottom, and additional connections between sectors to suppress the excitation of the elastic bending modes of the metaatom. 
\begin{figure}[t!]
\centering
\includegraphics[width=1.0\linewidth]{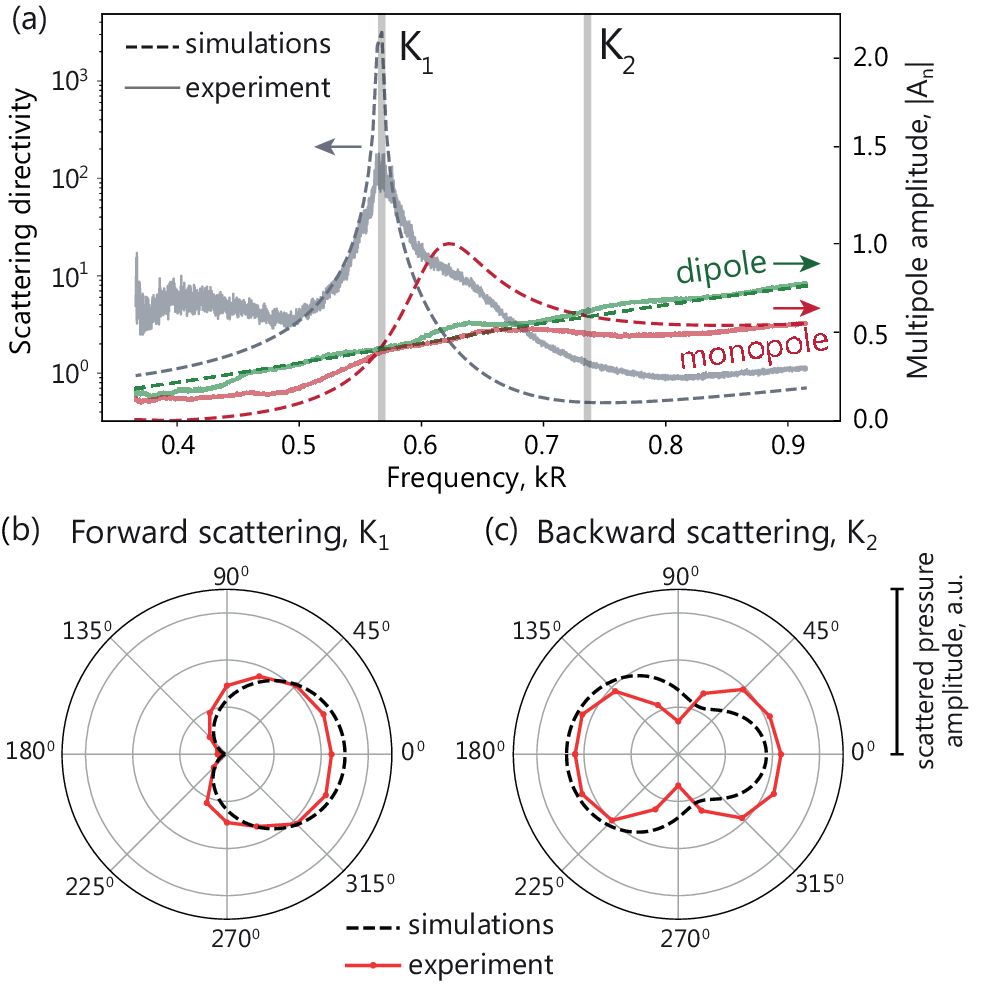}
\caption{(a) Directivity, defined as the ratio between forward and backward scattering intensities, obtained from simulation and experiment; (b,c) simulated and measured radiation patterns at the frequencies corresponding to the first and second Kerker conditions, i.e., maximum forward and backward scattering, respectively. The metaatom used in simulation and experiment has following geometrical parameters: $R=4$~cm, $r=1$~cm, $a=6.2$~mm, 5~sections, 3~bendings of channel.}
\label{fig3:Kerker}
\end{figure}
The sound source (Visaton BF 45/4 loudspeaker) was placed at one of the side of the chamber and at the considered distances the wave incident at the metaatom placed in the center can be effectively considered as a plane. Field detection is performed with a Crysound CRY371 microphone that can be moved to different angular positions around the metaatom. The phase information is obtained by referencing the microphone signal to a loopback from the source allowing for measurements of the absolute time delays between the emitted and collected signals. Systematic deviations in measured amplitudes and phases is compensated by carrying out reference measurements without the metaatom:  
\begin{equation}
    \bar{p}_{\text{sc}} = \frac{p_{\text{tot}} - p_{\text{b}}}{p_\text{b}},
\end{equation} 
where $p_{\text{tot}}$ and $p_{\text{b}}$ denote the total and background fields, respectively, and the division by background field is done to correct for amplitude and phase response of the measurement system. Here and further the scattered field is defined this way, both in the modelling and experiment.

The Fourier transformed time-domain signals collected in the points in front of and behind the metaatom, and their comparison to numerical modelling are presented in Fig~\ref{fig2:setup_data}(b) containing both amplitude and phase spectra. The measured data (solid lines) for the scattered field is in good agreement with the numerical results (dashed lines), however, noticeable discrepancy between numerical and experimental data is present in the quality of resonances. This is explained by the intrinsic thermoviscous losses in the metaatom channels and will be discussed further. 

Based on this data, the ratio of forward to backward scattering intensities was obtained, showing a pronounced peak of almost 1500 at the frequency of the first Kerker condition at the frequency $f=775$~Hz ($K_1$ point) as shown in Fig.~\ref{fig3:Kerker}(a). The far field angular diagram shown in Fig.~\ref{fig3:Kerker}(b) exhibits a typical cardioid shape perfectly matching with the simulated results. At the second Kerker condition ($K_2$ point, $1190$~Hz) the dominance of the backward scattering is not that pronounced and we see some discrepancy in the measured and simulated far-field diagrams Fig.~\ref{fig3:Kerker}(c). This can be attributed to experimental limitations, such as imperfect boundary absorption, that hinder the efficient excitation of high-quality resonances required for robust destructive interference, making it more challenging to achieve in the backward direction. Secondly, the acoustic losses stemming from viscous and thermal damping also destroy the resonances and introduce additional parasitic losses preventing the ideal interference condition.
\begin{figure}[t!]
\centering
\includegraphics[width=0.95\linewidth]{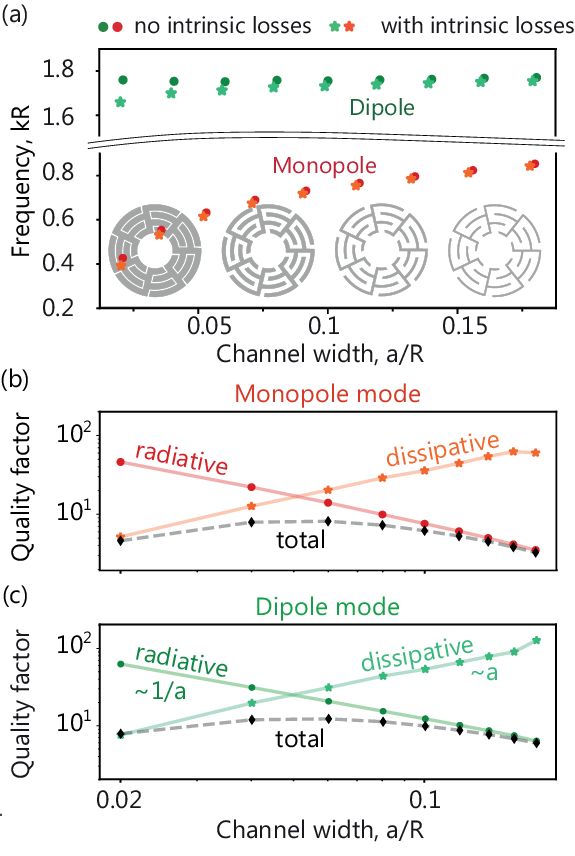}
\caption{(a) The dependence of resonance frequency of monopole and dipole modes 
on the channel width of the metaatom. Monopole (b) and dipole (c) mode intrincic, $Q_{\mathrm{rad}}$, radiative, $Q_{\mathrm{diss}}$ and total, $Q_{\mathrm{tot}}$, loss quality factor dependence on the channel width of the metaatom.}
\label{fig:losses}
\end{figure}

The effect of thermoviscous losses deserves additional attention as it not discussed in details in the literature in the view of labyrinth metaatoms. Due to its resonant properties, the metaatom confines the the field inside narrow labyrinth channels, which inevitably leads to high viscous losses~\cite{jordaan2018}. To quantify the radiative and intrinsic quality factors of the acoustic resonator modes, we performed two eigenfrequency analyses in COMSOL Multiphysics. First, the radiation-only complex eigenfrequency $\omega_{\mathrm{rad}}$ was obtained using the \textit{Pressure Acoustics} module, which neglects viscous and thermal boundary-layer losses. Second, the full complex eigenfrequency $\omega_{\mathrm{tot}}$ was calculated by implementing \textit{Thermoviscous Boundary Layer Impedance} on the walls of metaatom in \textit{Pressure Acoustics} module to account for viscous and thermal dissipation within the acoustic boundary layers. The imaginary parts of these eigenfrequencies yield the corresponding modal decay rates. The quality factors can then be computed as:
\begin{align}
Q_{\mathrm{rad}} = \frac{\mathrm{Re}\{\omega_{\mathrm{rad}} \} }{2  \,\left|\mathrm{Im}  \{ \omega_{\mathrm{rad}} \} \right| }, \quad Q_{\mathrm{diss}} = \frac{\mathrm{Re}\{\omega_{\mathrm{tot}} \} }{2 \left|  \mathrm{Im}  \{ \omega_{\mathrm{tot}} \} - \mathrm{Im}  \{ \omega_{\mathrm{rad}} \} \right| }.
\end{align}

In order to exploit the resonant effects of the metaatom, the condition $Q_{\mathrm{rad}} \ll Q_{\mathrm{diss}}$ must be satisfied. The final geometry of the metaatom was optimized by analysing the dependence of resonant frequency and the ratio of radiative and intrinsic quality factors $Q_{\mathrm{rad}}/Q_{\mathrm{diss}}$ on the channell width, depicted in Fig~\ref{fig:losses}. It can be seen that the shift in resonance frequency is small (see Fig.~\ref{fig:losses}(a). Meanwhile the quality factors for the intrinsic and radiative losses vary significantly with change in metaatom channel width, see Fig.~\ref{fig:losses}~(b,c). Particularly, intrinsic quality factor rises linearly in~$ka$, while the radiative quality factor is inversely proportional to $ka$. This behaviour can also be recovered analytically, as the losses in labyrinthine metaatoms can be well approximated by losses in acoustic ducts of equivalent effective length~\cite{Chen2024}. The intrinsic and radiative quality factors of an open resonator supporting plane standing waves can equivalently be calculated as~\cite{kinsler}:
\begin{align}
    Q_{\mathrm{rad}(\mathrm{diss})} = \frac{k}{2\alpha_{\mathrm{rad}(\mathrm{diss)}}},
\end{align}
where $k$ is the real part of wavenumber, and $\alpha_{\mathrm{rad}(\mathrm{diss})}$ is the attenuation constant stemming from radiation (intrinsic loss). 

The attenuation constant from radiation for a pipe with one open end is of the form~\cite{isakovich} $\alpha_{\mathrm{rad}} \propto k\,\mathrm{Re}\left[Z_{\mathrm{rad}}\right]$, where $\mathrm{Re}\left[Z_{\mathrm{rad}}\right]$ is radiation resistance of the open end. Assuming $ka \ll 1$, the opening can be modelled as a radiating piston, the radiation resistance of which in two dimensions, for small ka, is~\cite{Mellow2011} $\mathrm{Re}\left[Z_{\mathrm{rad}}\right] \propto ka$. This gives overall dependence of $Q_{\mathrm{rad}} \propto 1/a$. 

Identically we can deduce the dependence of intrinsic factor if we know the dependence of $\alpha_{\mathrm{diss}}$ on the width of the channel. For two and three dimensional cases~\cite{kinsler,Cotterill2018}, the dependence is $\alpha_{\mathrm{diss}} \propto\sqrt{k}/a$, so the dependence of intrinsic Q factor is $Q_{\mathrm{diss}}\propto\sqrt{k}a$. 

Both $Q_{\mathrm{rad}}$ and $Q_{\mathrm{diss}}$ align with the numerical results in Fig~\ref{fig:losses}. Such nature of their dependence means that losses can be tuned by changing the channel width to better observe the desired effect, which was performed and the finalized metaatom geometry was chosen.

In summary, we have reported the first experimental observation of the acoustic analogue of the Kerker effect using labyrinthine resonators in a two-dimensional waveguide environment. By tailoring monopolar and dipolar resonances of a coiled-space metaatom, we demonstrated forward directional scattering in excellent agreement with theoretical predictions, thereby validating the feasibility of multipole interference for sound control. The backward Kerker condition was found to be less pronounced, primarily due to intrinsic thermoviscous losses, which limit the quality of resonant interference. Nevertheless, the presented results establish a clear acoustic counterpart of the Kerker effect and provide a foundation for future developments in directional acoustic devices.

\begin{acknowledgments}
The work is funded by Russian Science Foundation grant No. 25-79-31027, https://rscf.ru/project/25-79-31027/.
\end{acknowledgments}

\bibliography{apssamp}

\end{document}